\documentclass[final,1p,times]{elsarticle}

\usepackage{graphicx}
\usepackage{amssymb}
\usepackage{amsmath}

\usepackage{color}

\begin{document}

\begin{frontmatter}


\title{Multiplex networks of musical artists: the effect of heterogeneous inter-layer links}



\author[a,b,c]{Johann H. Mart\'inez}
\author[d]{Stefano Boccaletti}
\author[e]{Vladimir V. Makarov}
\author[b,c,f]{Javier M. Buld\'u  \corref{mycorrespondingauthor}}
\cortext[mycorrespondingauthor]{Corresponding author}
\ead{javier.buldu@urjc.es}

\address[a]{INSERM-Institute du Cerveau et de la Moelle \'Epini\`ere. H. Salp\^etri\`ere. Paris. France}
\address[b]{Grupo Interdisciplinar de Sistemas Complejos (GISC). Madrid, Spain}
\address[c]{Laboratory of Biological Networks, Center for Biomedical Technology - UPM. Madrid, Spain}
\address[d]{CNR-Institute of Complex Systems. Florence, Italy}
\address[e]{Yuri Gagarin State Technical University of Saratov, Saratov, Russian Federation}
\address[f]{Complex Systems Group, Universidad Rey Juan Carlos. Madrid, Spain}

\begin{abstract}
The way the topological structure goes from a decoupled state into a coupled one in multiplex networks has been widely studied by means of analytical and numerical studies, involving models of artificial networks. In general, these experiments assume uniform interconnections between layers offering, on the one hand, an analytical treatment of the structural properties of multiplex networks but, on the other hand, loosing applicability to real networks where heterogeneity of the links' weights is an intrinsic feature. In this paper, we study 2-layer multiplex networks of musicians whose layers correspond to empirical datasets containing, and linking the information of: (i) collaboration between them and (ii) musical similarities. In our model, connections between the collaboration and similarity layers exist, but they are not ubiquitous for all nodes. Specifically, inter-layer links are created (and weighted) based on structural resemblances between the neighborhood of an artist, taking into account the level of interaction at each layer. Next, we evaluate the effect that the heterogeneity of the weights of the inter-layer links has on the structural properties of the whole network, namely the second smallest eigenvalue of the Laplacian matrix (algebraic connectivity). Our results show a transition in the value of the algebraic connectivity that is far from classical theoretical predictions where the weight of the inter-layer links is considered to be homogeneous. \end{abstract}


\end{frontmatter}



\section{Introduction}

{\it Network Science} relies on four fundamental pillars: graph theory, statistical physics, nonlinear dynamics and Big Data \cite{newman2010}.
During the Internet era, the access to large amounts of datasets has led to the publication of a tremendous number of scientific articles analyzing real datasets, many of them using methodologies and tools coming from network science. The analysis of social systems has been one of the fields that has benefited the most and, specifically, how the particular patterns of interaction between individuals constrain a diversity of processes occurring in society, from opinion formation to disease spreading \cite{fortunato2009}. In this way, understanding the particular features reported in social networks, such us the small-world organization, the existence of communities or the high heterogeneity of the number of interactions between individuals, have been the keys for identifying some of the underlying rules behind the organization and dynamics of social networks.
However, during the last years, the development of new methodologies to construct and analyze multilayer networks has given a second breath to the field of network science \cite{kivela2014, boccaletti2014}. Multilayer networks are, in fact, networks composed of several layers that are inter-connected between them, each layer containing, in general, a specific kind of interaction between nodes.
Multiplex networks are a particular kind of multilayer networks where links between layers only connect the projections of the same node at the different layers \cite{kivela2014}. For example, consider a Facebook-Twitter multilayer network, where layer A contains the Facebook friendship of a series of people and layer B contains the interactions in Twitter (e.g., likes, retweets or messages) of the same group of people. Without loss of generality, let's asume that both layers contain the same number of nodes, i.e. user accounts, despite it is not a necessary condition for multiplex networks, and in our particular example, a person has an account at both social networks. Note that connections in layer A are not necessary the same connections as in layer B, since one may interact with a certain colleague in Facebook and not in Twitter, or vice-versa. Finally, inter-layer links are created between the two accounts of the same person, leading to a 2-layer multiplex network.

One crucial issue of the analysis of multiplex networks is the definition of the inter-layer links. While interactions inside a layer are defined by the nature of the layer itself (e.g., in the example of Facebook-Twitter multiplex networks, could be defined as the interactions between users at each of the online platforms), the way a node transmits its state from one layer to another is, at least, difficult to be evaluated. Taking our previous example, it would account for the probability of transmitting an information acquired in Facebook, to Twitter and conversely.
To overcome this issue, the most extended strategy is to assign a parameter $p$ to the weight of the inter-layer links and evaluate the consequences of modifying its value on the topological properties of the multiplex network \cite{radichi2013}. Sweeping the value of $p$, it is possible to analyze the importance of inter-layer links and to detect different transitions at the spectral properties of the multiplex networks \cite{radichi2013,gardenes2013} or even using it for classifying purposes, as in the case of multiplex functional networks of patients suffering from epilepsy \cite{dedomenico2016}. However, recent results have shown that, in real systems, the weight of the inter-layer links is not necessary homogeneous, and that the diversity of weights can induce important differences in the network properties \cite{buldu2018,makarov2016}.

In the current paper, we investigate the effects of the heterogeneity of the inter-layer links on the structure of multiplex networks of musical artists.
Since the beginning of twenty-first century and due to the recent ability to access large (on-line) datasets, music networks have captured the attention of a diversity of scientists trying to analyze how music collaboration, similarity and diversity spread along the social network formed from musical interests, no matter if the fundamental nodes where music consumers or musical artists.
For example, in Cano et al. \cite{cano2006}, the structure of four different on-line platforms for music recommendation were inspected, unveiling two different kinds of strategies for recommending music, one based on popularity and the other one based on music similarity. Based on coincidence of musical hits in personal playlists, Buld\' u et al. \cite{buldu2007} constructed a network of musical tastes and analyzed the evolution of their properties along time, which allows identifying the topological properties of top-hits.
Other studies have focused on the understanding of how collaboration between artists arises and its influence in music similarity. For example, in \cite{teitelbaum2008} community detection algorithms were used to identify the role of musical leaders both in collaboration and similarity networks, distinguishing between local leaders, whose influence was restricted to a specific musical genre and connector leaders linking different musical styles. More recently, Park et al. \cite{park2015} investigated the co-occurence of classical music composers in CDs (compact discs) with the aim of understanding the centrality of western classical composers. Interestingly, a superlinear preferential attachment was found as the explanation for the increasing concentration of edges around top-degree nodes and the arousal of power-law degree distributions, which allowed authors to  forecast the future of several prominent composers \cite{park2015}.

As in \cite{park2006}, we are interested in the interplay between collaboration and similarity networks of musical artists. However, our objective is not to compare the
topologies of both networks but to integrate the information contained in them. With this aim, we constructed a 2-layer multiplex network composed of a (i) collaboration and a (ii) similarity layer. In this way, each artist is represented by a node at each layer, which is connected to the corresponding neighbours according to (i) having collaborated with him/her and (ii) play similar music. In the process of obtaining the multiplex construct, the most crucial point is the creation of links connecting the representations of an artist at each layer. We propose a model based on neighbour's resemblance to quantify the weight of the inter-layer links and to analyze the differences with the theoretical case where all inter-layer links have the same weight. As we will see, our results show that the heterogeneity of the inter-layer links has important consequences on the spectral properties of the multiplex networks, which suggests that the theoretical predictions obtained with the homogenous approximation must be taken as limit cases but not as an accurate description of real multiplex networks.

\section{Materials and Methods}
The dataset consists on a curated version of the collaboration and similarity networks analyzed in \cite{park2006}, which were obtained from the AllMusicGuide web site \cite{allmusic}.
Specifically, we gathered artists' meta-data to built networks based on two types of edges: Collaboration ($C$) and Similarity ($S$). The former is the category in which two musicians are linked if they have played together in one or more albums, meanwhile the latter is the one in which two artists are linked if they play similar music according to the musical criteria of the AllMusicGuide musical editors \cite{cano2006,park2006}. For instance, the $3^{th}$ single of \textit{The Marshall Mathers} LP released in 2000 was played by a rap musician and a pop singer. Since both artists belong to different music styles they are joined in the collaboration network $C$, but not in the similarity network $S$.
\begin{figure}[h!]
    \centering
    \includegraphics[width=\textwidth]{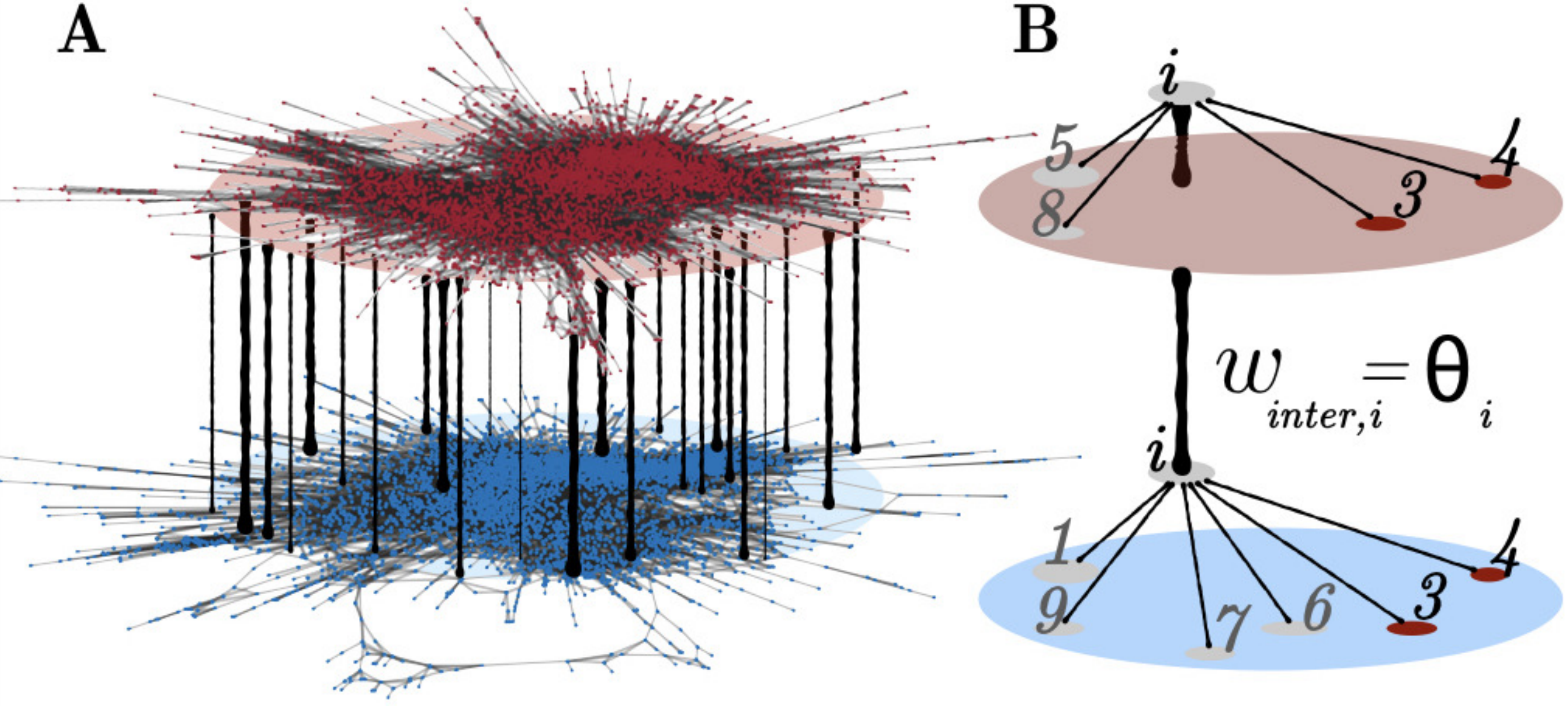}
    \caption{\textbf{A}. Multiplex representation of $C$ (red) and $S$ (blue) networks with heterogeneous weights at the inter-layer links. Note that inter-layer links have different thickness (weights) and not all of them are present.
    \textbf{B}. Qualitative example of the overlapping method: For an artist $i$, we pick up its common neighbours in both layers $O_i^{C,S}=2$ (in this example, artists 3 and 4) and its maximum degree $max(k_i^C,k_i^S)=6$ in order to get $\theta^{'}_i=1/3$. The final weight of the inter-layer links of an artist $i$ is $w_{inter,i}=\theta_i=\theta^{'}_i/\langle\{ \theta^{'}_i \}\rangle$, with $\{ \theta_i \}$ having an average value of one, by definition.}\label{fig:01}
\end{figure}

In order to guarantee that both networks do not have isolated components, first, we selected all artists that belong to the giant component of either network $C$ or $S$ and, second, we kept only those artists that appeared at both networks (maintaining the connectivity of the giant component). In this way, we obtain a total number of $N = 4673$ artists (i.e., nodes), with $L_C=16150$ links at the collaboration network and $L_S=15830$ at the similarity network. Note that both $L_C$ and $L_S$ refer to intra-layer links.
In contrast to the classical multiplex model where all inter-layer links have the same value, we built a more actual representation by assembling both layers (networks) with a set of \textit{ad-hoc} inter-layer links endowed with characteristics of both networks.
Next, we construct the multiplex network of musical artist by connecting the two nodes representing the same artists at each layer. We quantified the weights of the inter-layer links by means of an overlapping method, which takes into account the overlap between the local neighborhood of an artist $i$ in both layers.
In this way, we identified the node representing artist $i$ at both layers $C$ and $S$ and we count the number of common neighbours $O_i^{C,S}$, as well as the maximum degree of the artist at both layers $max(k_i^C, k_i^S)$. We obtained the overlapping parameter $\theta^{'}_i=O_i^{C,S}/max(k_i^C, k_i^S)$ for each of the $N$ artists of the multiplex network. Next, we defined the normalized overlapping parameter $\theta_i=\theta^{'}_i/\langle\{ \theta^{'}_i \}\rangle$, which guarantees that the average value of $\theta_i$ is equal to one (see Fig. \ref{fig:01}, for a qualitative description). Finally, we assigned a weight $w_{inter,i}=\theta_i$ to the inter-layer link of artist $i$. Note that, in this way, we are evaluating the amount of interaction between similarity and collaboration for a given artist: if the structure of the neighborhood is similar in both layers/networks, we are assuming that they are somehow influencing each other, while low values of $\theta_i$ indicate that collaborations and similarity of a given artist are quite independent.

\section{Results}
Here, we are interested about the consequences of using a classical multiplex approach with all inter-layer links being equal or, conversely, considering a non-homogeneous distribution of inter-layer weights.
With this aim, we are going to focus on the impact of both approaches on the algebraic connectivity $\lambda_2$ of the multiplex artist network. The algebraic connectivity corresponds to the second eigenvalue of the supra-Laplacian matrix $\mathcal{L}$, which is obtained as  $\mathcal{L}=D-M$, where $D$ is a diagonal matrix containing the degree of each node and $M$ is the supra-adjacency matrix of the multiplex network:

\begin{equation}
M=
\begin{pmatrix}
    C      & p\mathbf{I}  \\
    p\mathbf{I}      & S \\
\end{pmatrix}
\end{equation}

where $\mathbf{I}$ is the $N\times N$ identity matrix, $C$ and $S$ are the adjacency matrices of the collaboration and similarity layers and $p$ is a parameter containing the weight of the inter-layer links, which is commonly considered as a control parameter of the interaction between layers \cite{radichi2013}. The algebraic connectivity is given by  $\lambda_2$, which is the smallest non-zero eigenvalue of the Laplacian matrix where $\lambda_1=0 \leq \lambda_2 \leq \lambda_3 = ... = \lambda_N$ (note that $\lambda_1=0$ always, since the Laplacian matrix fulfills, by definition, the zero-row sum condition).

We are concerned about the algebraic connectivity since it is a parameter that gives useful information about the diffusion properties of the network \cite{gomez2013}, its ability to synchronize \cite{aguirre2014} and its modularity \cite{newman2006}.

\begin{figure}[h!]
    \centering
    \includegraphics[width=\textwidth]{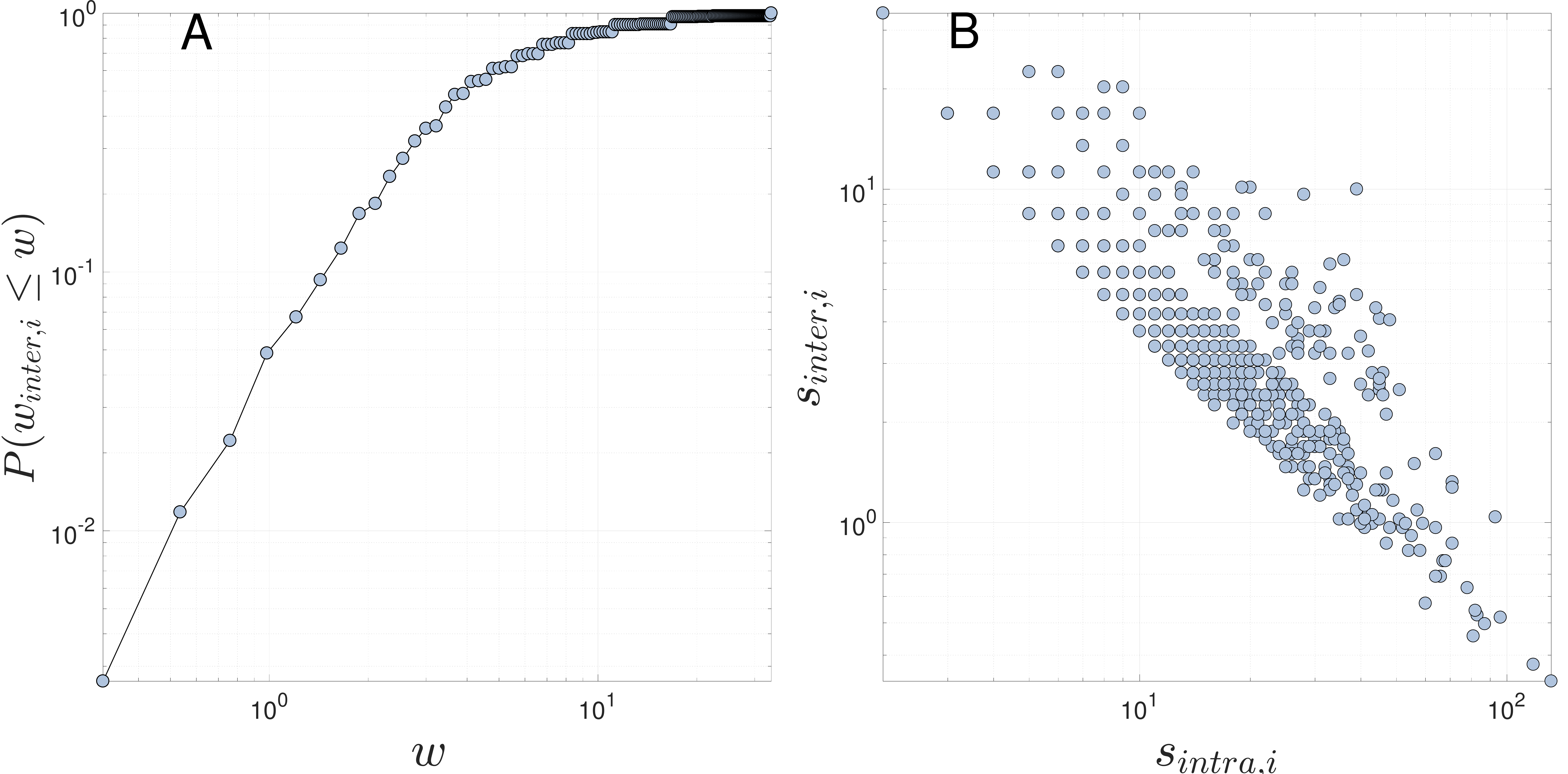}
    \caption{\textbf{A}. Cumulative distribution function (CDF) of the inter-layer weights $w_{inter,i}=\theta_i$. Specifically, the distribution accounts for the probability of finding a weight  $w_{inter,i}$ equal or lower than a value $w$. Inter-layer links with $w_{inter,i}=0$ have been excluded from the probability distribution.
    \textbf{B}. Interdependency between inter-layer strength $s_{inter}$ of a musical artist and his/her total intra-layer strength $s_{intra}$. Note the negative correlation.}\label{fig:02}
\end{figure}

First, in order to evaluate the heterogeneity of the inter-layer links obtained with the overlapping method, we calculate the cumulative distribution function (CDF) of  $\{w_{inter,i} | w_{inter,i}\neq0\}$ with $w_{inter,i}=\theta_i$, which is plotted in Fig. \ref{fig:02}A. Note that nodes without common neighbours at both layers have $w_{inter,i}=0$ ($\sim$ 83\%), while the remaining ones are quite heterogeneous, since the values span over two orders of magnitude.

Next, we appraised the dependency of the inter-layer strength of the nodes at each layer with the individual intra-connectivity inside each layer. Specifically, we define the inter-strength $s_{inter,i}$ of an artist $i$ as the inter-layer weight $\{w_{inter,i}\}$ (note that, since we are dealing with multiplex networks, there is only one inter-layer link per artist), while the artist intra-layer strength is obtained as $s_{intra,i}=k^C_i+k^S_i$, i.e. the sum of the neighbors of an artist at the two layers. Figure \ref{fig:02}B shows the negative trend existing in the relationship between $s_{inter,i}$ and $s_{intra,i}$. As a consequence, the most connected nodes of each layer (i.e., the layer hubs) have, in general, a low $\theta_i$, due to the lack of common neighbours at both layers. On the contrary, low-degree nodes are prone to have a higher overlap, rendering a higher $\theta_i$. This fact shows that the robustness of our multiplex scaffolds seems to lie on peripheral nodes. Thus, it naturally raises the question of how the properties of the multiplex network may change with this type of inter-layer heterogeneity. To answer this question, we compared the algebraic connectivity of both multiplex models, i.e., the classical one with the inter-layer links having equal weights, and the one using the overlapping method.

Note that, the algebraic connectivity $\lambda_2$ of the multiplex network is an indicator of how independent layers $C$ and $S$ are. As shown by Radichi and Arenas \cite{radichi2013}, when the inter-layer links have the same value $w_{inter,i}=p$, it is possible to detect two different structural regimes when the value of the parameter $p$ is increased from zero.
For low to moderate values of $p$, $\lambda_2$ grows independently of the internal structure of both layers and it is given by the linear function $\lambda_2=2 \times p$. However, $\lambda_2$ suffers an abrupt change when crossing a certain threshold $p>p_c$ and enters into what is called the coupled phase. In this regime ($p\geq p_c$) $\lambda_2$ is governed by a monotonically increasing function that saturates at $\lambda_2= \lambda_2 \{\mathcal{L}_C+\mathcal{L}_S\}/2$, with $\{\mathcal{L}_C+\mathcal{L}_S\}$ being the Laplacian matrix of the aggregated network, consisting in collapsing layers $C$ and $S$ into a single one of size $N$, just by summing the links of both layers.
Therefore, this latter region directly depends on the structure of both layers, a fact that did not happen for values of $p$ below $p_c$.
Importantly, the critical parameter $p_c$, defining the border between both regimes, can be obtained by inspecting the evolution of $\lambda_3$ (the second non-zero eigenvalue of the Laplacian matrix) and detecting which value of $p$ leads to a crossing with $\lambda_2$.

\begin{figure}[h!]
    \centering
    \includegraphics[width=\textwidth]{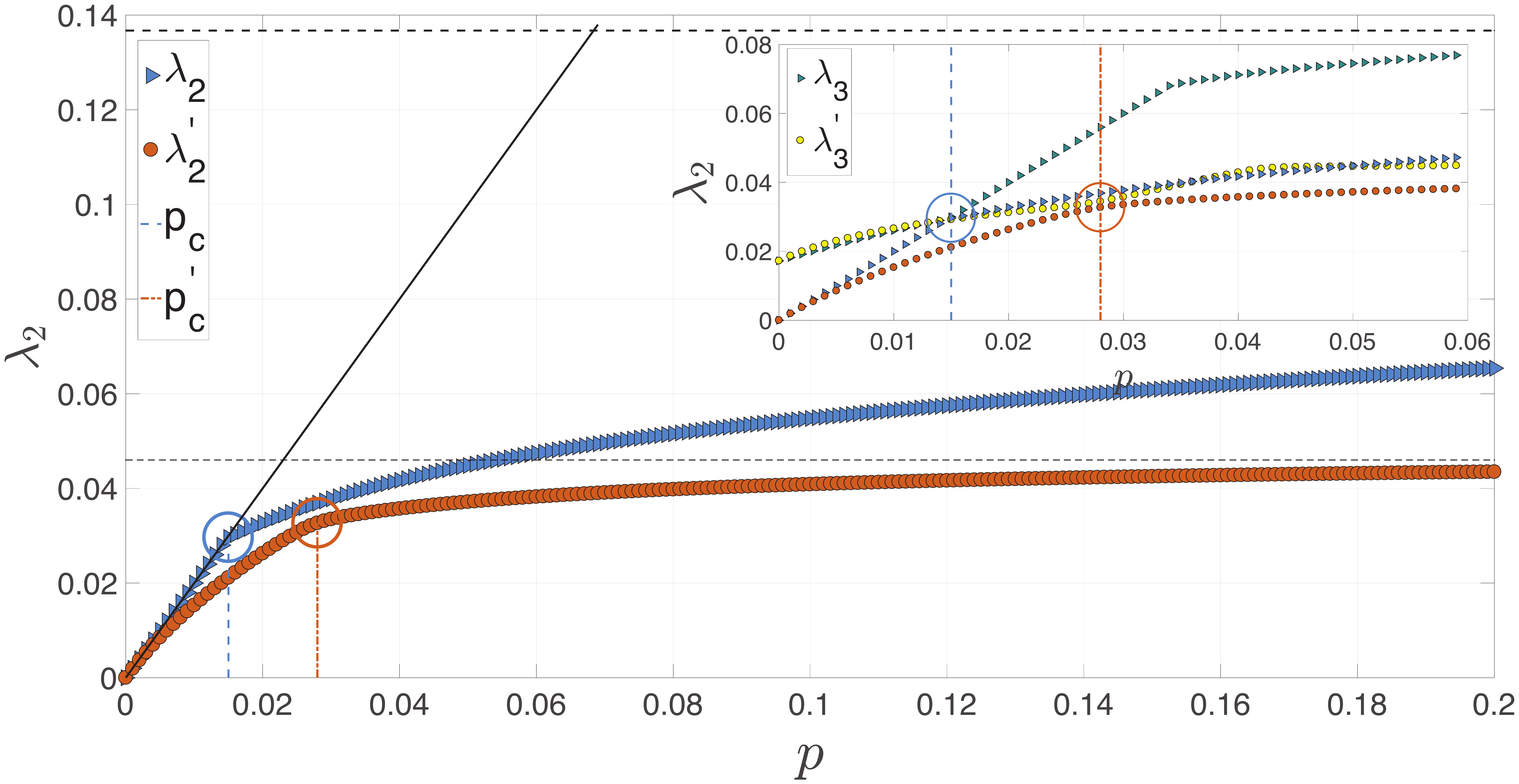}
    \caption{Algebraic connectivities vs the control parameter $p$, which directly controls the average weight of the inter-layer links. Black solid line corresponds to $\lambda_2= 2 \times p$ and the horizontal dashed line is algebraic connectivity of the aggregate network $\lambda_2= \lambda_2 \{\mathcal{L}_C+\mathcal{L}_S\}/2$.  Blue triangles are the values of $\lambda_2$ for the classical model (all inter-layer weights are $p$), while red circles correspond to the case of heterogeneous inter-layer links, where inter-layer weights are given by $w_{inter,i}=p \times \theta_i$. The inset is a zoom of the main figure where $\lambda_3$ (green triangles) and $\lambda_3^{'}$ (yellow circles) have been included to ease the observation of $p_c$ and $p_c^{'}$ (the point where $\lambda_2$ and $\lambda_3$ cross each other).}\label{fig:03}
\end{figure}


Figure \ref{fig:03} shows the values of $\lambda_2$ (homogeneus inter-layer links) and $\lambda_2^{'}$ (heterogeneous inter-layer links) depending on the amount of coupling between layers. Specifically, $\lambda_2$ is obtained by setting the weights of all inter-layer
links to $w_i=p$, with $p$ being modified from zero to 0.2 in steps of $p=0.001$. Similarly, $\lambda_2^{'}$ corresponds to the algebraic connectivity of the heterogeneous case for increasing values of $p$, in this case with the inter-layer links given by $w_i=p \times \theta_i$. Note that, since $\langle\{ \theta_i \}\rangle=1$, the sum of the weights of the inter-layer links is the same in both models for the same value of $p$. In Fig. \ref{fig:03}, we also indicated the position of the thresholds  $p_c$ (homogeneous case) and $p_c^{'}$ (heterogeneous case), which are identified in the
figure inset through the crossing of
$\lambda_2$ and $\lambda_3$. The second and third lowest eigenvalues of the Laplacian matrix approximate asymptotically, but never touch each other and the minimum distance between them gives the coordinates of $p_c$ (and $p_c^{'}$ for $\lambda_2^{'}$ and $\lambda_3^{'}$). Interestingly, the transition point of the real multiplex network is delayed by the heterogeneity of the inter-layer links and because the high percent of them that are zero. On the other hand, $\lambda_2^{'}$ rapidly saturates around $0.046$, a value almost three times lower than the saturation level of $\lambda_2$  ($0.1367$), which corresponds to the algebraic connectivity of the aggregate network.

The latter fact opens the door to questioning about whether the gap between saturation levels remains constant or is altered under perturbations of one of the layers. To analyze this issue, we have modified the weight of the connections inside a layer and studied the consequences on $\lambda_2$ and $\lambda_2^{'}$. The importance of one network over the other depends on the ratio of the total layer strength (i.e., the ratio of the sum of the weights of all links of each layer). Thus, we have taken the collaboration network $C$ and multiplied the values of the weights of its intra-layer links by a parameter $\gamma$, which is modified from $0.2$ to $2$. In this way, we are decreasing the overall strength of the collaboration layer for $\gamma < 1$ and increasing it for $\gamma_1>1$, which results on decreasing/increasing the importance of collaboration vs similarity in the multiplex artist network.


\begin{figure}[h!]
    \centering
    \includegraphics[width=\textwidth]{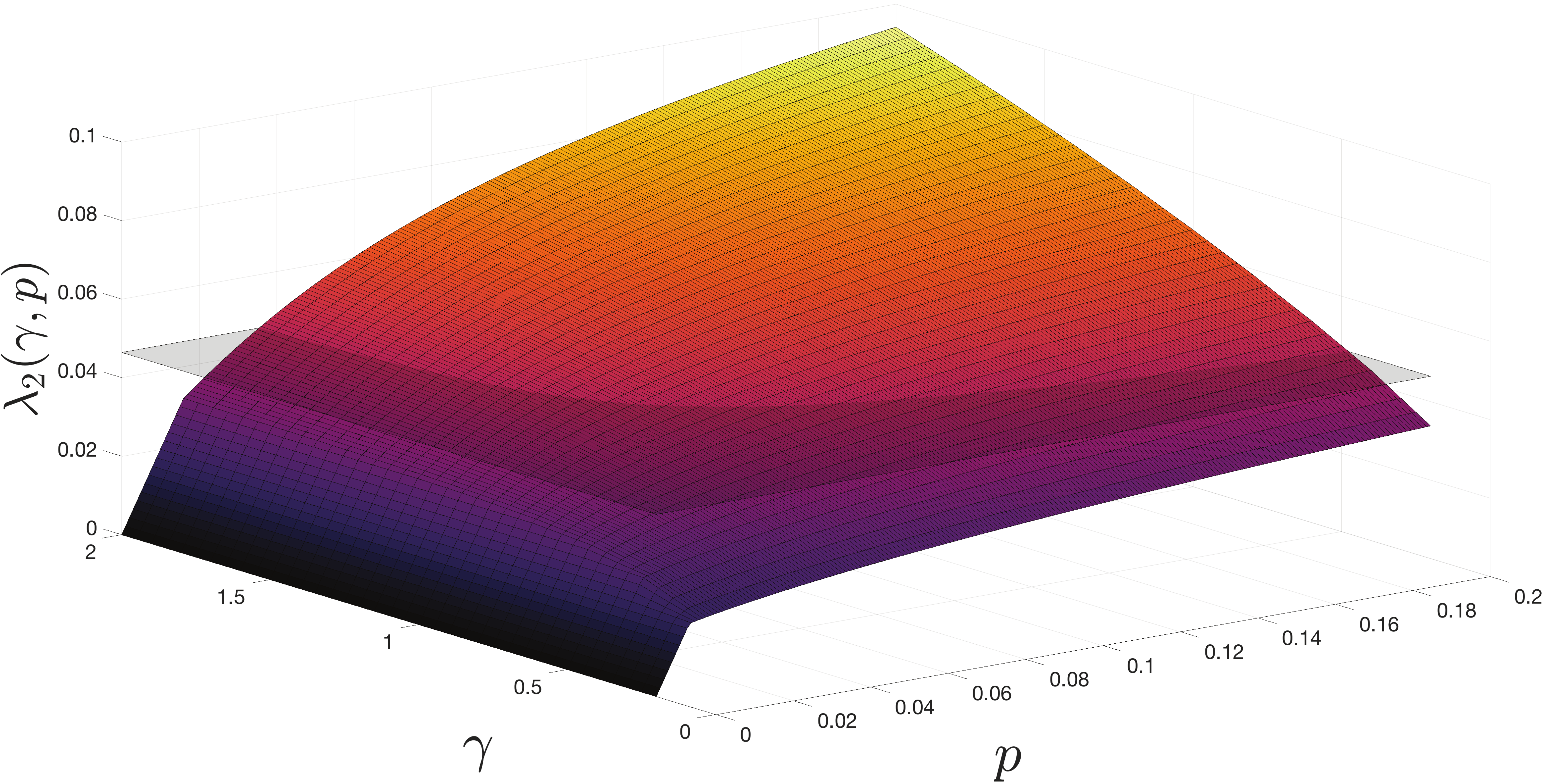}
    \caption{Algebraic connectivity $\lambda_2(\gamma,p)$ for the perturbed layer $C(\gamma)$, with $0.2 \leq \gamma \leq 2.0$ and  $0 \leq p \leq 0.2$.
    In the classical model, each $C(\gamma)$ version leads to a saturation value given by  $\lambda_2 = \lambda_2 \{\mathcal{L}_{C(\gamma)}+\mathcal{L}_S\}/2$ (not shown here). Note that the higher the perturbation $\gamma$, the higher the $\lambda_2(\gamma,p)$ is.Gray $\gamma$-$p$ plane represents the value of $\lambda_2^{'}$ at which the heterogeneous case saturates (see Fig. \ref{fig:05} for details).}\label{fig:04}
\end{figure}

\begin{figure}[h!]
    \centering
    \includegraphics[width=\textwidth]{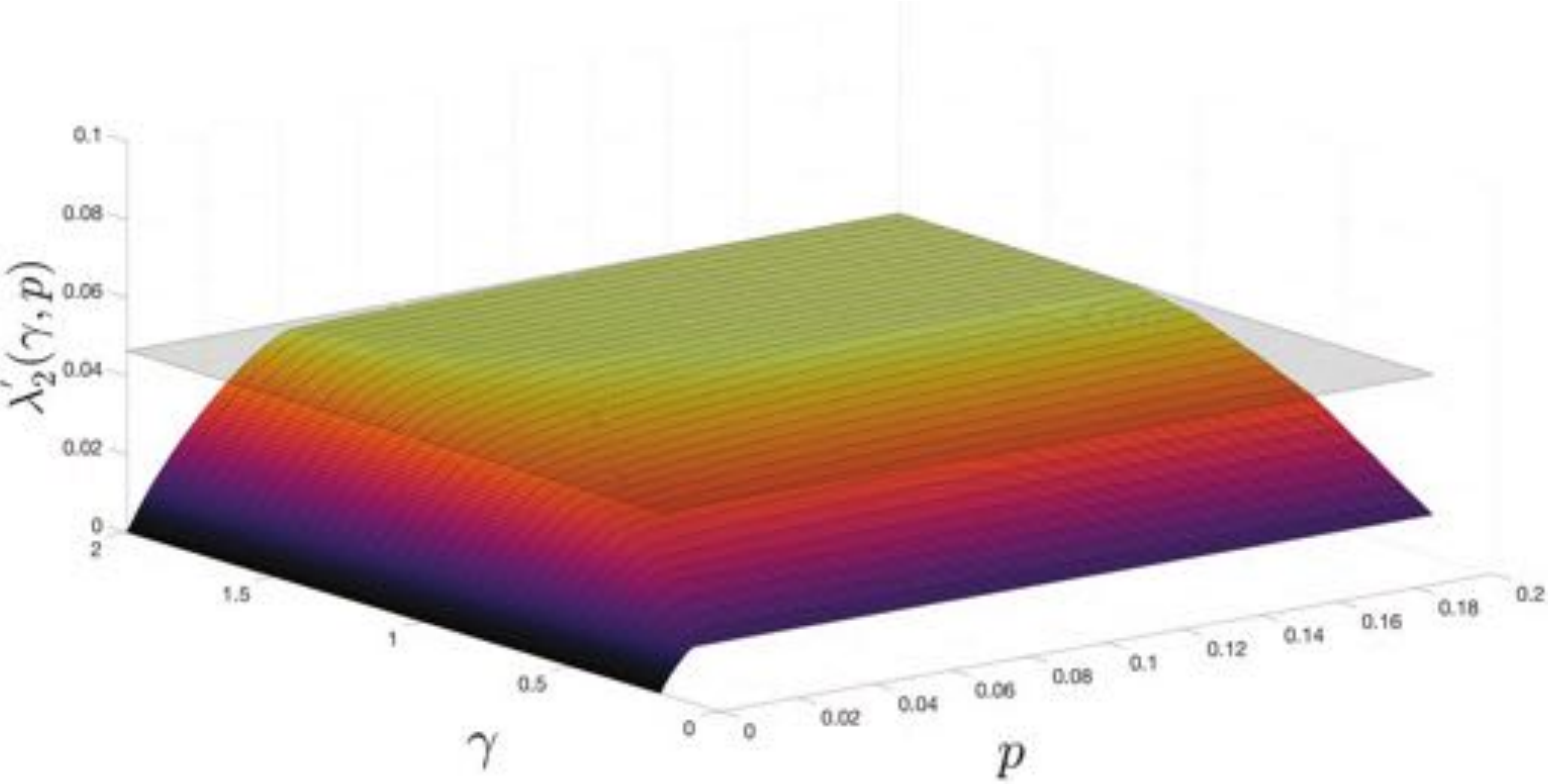}
    \caption{Algebraic connectivity $\lambda_2^{'} (\gamma,p)$ of the heterogeneous case for the perturbed layer $C(\gamma)$, with $0.2 \leq \gamma \leq 2.0$  and  $0 \leq p \leq 0.2$. Despite the increase of $\gamma$, $\lambda_2^{'} (\gamma,p)$ saturates at the value indicated by the gray $\gamma$-$p$ plane.}\label{fig:05}
\end{figure}

Figure \ref{fig:04} shows the impact of modifying the strength of $C$ on the algebraic connectivity $\lambda_2(\gamma,p)$ in case of having the same weight $w_{inter,i}=p$ for all inter-layer links. We observe how the algebraic connectivity continuously grows as the perturbation level $\gamma$ increases, which is basically a consequence of increasing the average strength of the layer. However, the behaviour of algebraic connectivity changes when we perform the same experiment using the heterogeneous distribution. As it can be seen in Fig. \ref{fig:05}, for heterogeneous inter-layer coupling, when $\gamma$ reaches values higher than one, $\lambda_2(\gamma,p)$ saturates around 0.0464, which is only $\sim$ 40 \% of the saturation value in the homogeneous model.
This fact indicates that differences between the homogeneous and heterogeneous distribution of inter-layer weights are (i) highly dependent on the differences of strength between
the layers of the multiplex network and (ii) have a non-linear behavior, with a saturation region far away from the analytical predictions given by the homogeneous model.

\section{Conclusions}
We investigated the effects of the heterogeneity of inter-layer links on the spectral properties of multiplex networks, namely the algebraic connectivity given by the second smallest eigenvalue of the supra-Laplacian matrix. We were concerned about the impact on artist networks composed of two layers, one containing collaborations between musical artists and the other one accounting for musical similarities. We defined an overlapping method to quantify the weight of the inter-layer links, which takes into account the similarity of the local neighbourhood of an artist at each layer. The distribution of the inter-layer links follows a heterogeneous distribution in contrast to the common uniform distribution in ideal multiplex representations. We investigated the transitions of the algebraic connectivity in both types of multiplex models (homogeneous and heterogeneous) and detected that the transition point of $\lambda_2$ is delayed in multiplex networks with heterogeneous weights. We also observed that $\lambda_2$  saturates in real multiplex networks compared to theoretical (homogeneous) ensembles. Finally, we showed that $\lambda_2$ saturates when one of the layer is perturbed by increasing the weight of its inter-layer links and that the saturation value is much lower than the one excepted in the homogeneous case. In sum, we have shown that a real representation of the multiplex significantly changes the  expected properties of theoretical multiplex networks, what should be taken into account in future studies with real datasets.

\section{Acknowledgments}
The authors acknowledge J. Aguirre, M. Chavez and P. L. del Barrio for fruitful conversations. J.M.B. is founded by MINECO (FIS2013-41057-P and FIS2017-84151-P). S.B. was supported by the project FIS2012-38949-C03-01 of MINECO. V.V.M. was supported by the Ministry of Education and Science of Russia (project 3.861.2017/4.6). Datasets were acquired thanks to the financial support of project FIS2012-38949-C03-01 of MINECO.








\bibliographystyle{model1-num-names}







\end{document}